\documentclass[sigconf]{acmart}
\AtBeginDocument{%
  \providecommand\BibTeX{{%
    \normalfont B\kern-0.5em{\scshape i\kern-0.25em b}\kern-0.8em\TeX}}}

\usepackage{multirow}
\usepackage{tabularx}
\usepackage{color}
\usepackage{tabularray}
\usepackage{xcolor}	
\usepackage{url}	
\usepackage{xspace}	
\usepackage{caption}	
\usepackage{subcaption}	
\usepackage{float}	
\usepackage{hyperref}	
\usepackage{cleveref}	
\usepackage{float}	
\usepackage{titlesec}
\usepackage{ragged2e}
\usepackage{booktabs}
\hypersetup{	
    colorlinks=true,	
    linkcolor=blue,	
    filecolor=magenta,      	
    urlcolor=cyan,	
}	

\usepackage{enumitem}
\setlist{leftmargin=3mm}

\copyrightyear{2024}
\acmYear{2024}
\setcopyright{rightsretained}
\acmConference[CHI '24]{Proceedings of the CHI Conference on Human Factors in Computing Systems}{May 11--16, 2024}{Honolulu, HI, USA}
\acmBooktitle{Proceedings of the CHI Conference on Human Factors in Computing Systems (CHI '24), May 11--16, 2024, Honolulu, HI, USA}
\acmDOI{10.1145/3613904.3642197}
\acmISBN{979-8-4007-0330-0/24/05}

\begin{document}

\definecolor{Author1}{HTML}{e41a1c}  
\definecolor{Author2}{HTML}{377eb8}  
\definecolor{Author3}{HTML}{4daf4a}  
\definecolor{Author4}{HTML}{984ea3}  
\definecolor{Author5}{HTML}{ff7f00}  
\definecolor{Reviewer}{HTML}{000000}  

\definecolor{Issue1}{HTML}{e41a1c}  
\definecolor{Issue2}{HTML}{377eb8}  
\definecolor{Issue3}{HTML}{4daf4a}  
\definecolor{Issue4}{HTML}{984ea3}  
\definecolor{Issue5}{HTML}{ff7f00}  

\newcommand{\reviewlegend}{
    \colorbox{Issue1}{\textcolor{white}{\#1}} \textcolor{Issue1}{Improving Presentation } \colorbox{Issue2}{\textcolor{white}{\#2}} \textcolor{Issue2}{Scoping to People with Autism } \colorbox{Issue3}{\textcolor{white}{\#3}} \textcolor{Issue3}{Deepening Implications for Design}
}

\newif\ifsubmit	

\submittrue	

\ifsubmit

\newcommand{\One}[1]{#1}	
\newcommand{\OneA}[1]{#1}	
\newcommand{\OneB}[1]{#1}	
\newcommand{\OneC}[1]{#1}	
\newcommand{\OneX}[1]{#1}

\newcommand{\Two}[1]{#1}	
\newcommand{\TwoA}[1]{#1}	
\newcommand{\TwoB}[1]{#1}	
\newcommand{\TwoC}[1]{#1}	
\newcommand{\TwoD}[1]{#1}	
\newcommand{\TwoE}[1]{#1}	
\newcommand{\TwoX}[1]{#1}

\newcommand{\Three}[1]{#1}	
\newcommand{\ThreeA}[1]{#1}	
\newcommand{\ThreeB}[1]{#1}	
\newcommand{\ThreeC}[1]{#1}	
\newcommand{\ThreeX}[1]{#1}

\newcommand{\Liz}[1]{#1}

\newcommand{\OneAC}[1]{}
\newcommand{\TwoAC}[1]{}
\newcommand{\RTwo}[1]{}
\newcommand{\RThree}[1]{}


\title[Collaborative Job Seeking for People with Autism]{Collaborative Job Seeking for People with Autism: Challenges and Design Opportunities}


\else
\newcommand{\zinat}[1]{\marginpar{\colorbox{Author1}{\textcolor{white}{ZA}} \textcolor{Author1}{#1}}}
\newcommand{\zinatIn}[1]{\colorbox{Author1}{\textcolor{white}{ZA}} \textcolor{Author1}{#1}}
\newcommand{\ray}[1]{\marginpar{\colorbox{Author2}{\textcolor{white}{RH}} \textcolor{Author2}{#1}}}
\newcommand{\rayIn}[1]{\colorbox{Author2}{\textcolor{white}{RH}} \textcolor{Author2}{#1}}
\newcommand{\vivian}[1]{\marginpar{\colorbox{Author3}{\textcolor{white}{VM}} \textcolor{Author3}{#1}}}
\newcommand{\vivianIn}[1]{\colorbox{Author3}{\textcolor{white}{VM}} \textcolor{Author3}{#1}}
\newcommand{\dasha}[1]{\marginpar{\colorbox{Author4}{\textcolor{white}{DP}} \textcolor{Author4}{#1}}}
\newcommand{\dashaIn}[1]{\colorbox{Author4}{\textcolor{white}{DP}} \textcolor{Author4}{#1}}
\newcommand{\slobodan}[1]{\marginpar{\colorbox{Author5}{\textcolor{white}{SV}} \textcolor{Author5}{#1}}}
\newcommand{\slobodanIn}[1]{\colorbox{Author5}{\textcolor{white}{SV}} \textcolor{Author5}{#1}}

\newcommand{\One}[1]{\colorbox{Issue1}{\textcolor{white}{1:Presentation}} \textcolor{Issue1}{#1}}
\newcommand{\OneA}[1]{\colorbox{Issue1}{\textcolor{white}{1A:Std1Std2Method}} \textcolor{Issue1}{#1}}
\newcommand{\OneB}[1]{\colorbox{Issue1}{\textcolor{white}{1B:Std1Std2Transparency}} \textcolor{Issue1}{#1}}
\newcommand{\OneC}[1]{\colorbox{Issue1}{\textcolor{white}{1C:Readability}} \textcolor{Issue1}{#1}}
\newcommand{\OneX}[1]{\textcolor{Issue1}{#1}}

\newcommand{\Two}[1]{\colorbox{Issue2}{\textcolor{white}{2:Scoping}} \textcolor{Issue2}{#1}}
\newcommand{\TwoA}[1]{\colorbox{Issue2}{\textcolor{white}{2A:LitScoping}} \textcolor{Issue2}{#1}}
\newcommand{\TwoB}[1]{\colorbox{Issue2}{\textcolor{white}{2B:Std1Scoping}} \textcolor{Issue2}{#1}}
\newcommand{\TwoC}[1]{\colorbox{Issue2}{\textcolor{white}{2C:DesignSpaceScoping}} \textcolor{Issue2}{#1}}
\newcommand{\TwoD}[1]{\colorbox{Issue2}{\textcolor{white}{2D:BridgingStd1.Std2}} \textcolor{Issue2}{#1}}
\newcommand{\TwoE}[1]{\colorbox{Issue2}{\textcolor{white}{2E:DeeperStd2}} \textcolor{Issue2}{#1}}
\newcommand{\TwoX}[1]{\textcolor{Issue2}{#1}}

\newcommand{\Three}[1]{\colorbox{Issue3}{\textcolor{white}{3:Implications}} \textcolor{Issue3}{#1}}
\newcommand{\ThreeA}[1]{\colorbox{Issue3}{\textcolor{white}{3A:DeeperDisc.}} \textcolor{Issue3}{#1}}
\newcommand{\ThreeB}[1]{\colorbox{Issue3}{\textcolor{white}{3B:AIImp.}} \textcolor{Issue3}{#1}}
\newcommand{\ThreeC}[1]{\colorbox{Issue3}{\textcolor{white}{3C:TransferableImp.}} \textcolor{Issue3}{#1}}
\newcommand{\ThreeX}[1]{\textcolor{Issue3}{#1}}

\newcommand{\OneAC}[1]{\marginpar{\colorbox{Reviewer}{\textcolor{white}{1AC}} \textcolor{Reviewer}{#1}}}
\newcommand{\TwoAC}[1]{\marginpar{\colorbox{Reviewer}{\textcolor{white}{2AC}} \textcolor{Reviewer}{#1}}}
\newcommand{\RTwo}[1]{\marginpar{\colorbox{Reviewer}{\textcolor{white}{R2}} \textcolor{Reviewer}{#1}}}
\newcommand{\RThree}[1]{\marginpar{\colorbox{Reviewer}{\textcolor{white}{R3}} \textcolor{Reviewer}{#1}}}

\title[\reviewlegend{}]{Collaborative Job Seeking for People with Autism: Challenges and Design Opportunities}

\author{Redacted for blind reviews, et al.}

\marginpar{\textbf{Legend}:
\begin{flushleft}
    \colorbox{Issue1}{\textcolor{white}{\#1}} \textcolor{Issue1}{Adding clarity and improving presentation:}\break
    \colorbox{Issue1}{\textcolor{white}{1A}} \textcolor{Issue1}{Std1 Method}\break
    \colorbox{Issue1}{\textcolor{white}{1B}} \textcolor{Issue1}{Std1 \& Std2 Materials}\break
    \colorbox{Issue1}{\textcolor{white}{1C}} \textcolor{Issue1}{Readibility} \break\break

    \colorbox{Issue2}{\textcolor{white}{\#2}} \textcolor{Issue2}{Improving the story and flow by more scoping to people with autism under job-seeking:}\break
    \colorbox{Issue2}{\textcolor{white}{2A}} \textcolor{Issue2}{Literature}\break
    \colorbox{Issue2}{\textcolor{white}{2B}} \textcolor{Issue2}{Std1 scoping} \break
    \colorbox{Issue2}{\textcolor{white}{2C}} \textcolor{Issue2}{Design Space scoping}\break
    \colorbox{Issue2}{\textcolor{white}{2D}} \textcolor{Issue2} {Connecting Std1 and Std2}\break
    \colorbox{Issue2}{\textcolor{white}{2E}} \textcolor{Issue2}{Deepening Std2}\break\break

    \colorbox{Issue3}{\textcolor{white}{\#3}} \textcolor{Issue3}{Improving study implications and discussion:}\break
    \colorbox{Issue3}{\textcolor{white}{3A}} \textcolor{Issue3}{Deepening Discussion}\break
    \colorbox{Issue3}{\textcolor{white}{3B}} \textcolor{Issue3}{Implications for AI}\break
    \colorbox{Issue3}{\textcolor{white}{3C}} \textcolor{Issue3}{Transferable implications} \break
\end{flushleft}
}


\fi




\author{Zinat Ara\textsuperscript{\textdagger}, Amrita Ganguly\textsuperscript{\textdagger}, Donna Peppard\textsuperscript{\S}, Dongjun Chung\textsuperscript{\textdaggerdbl}, Slobodan Vucetic\textsuperscript{\S}, Vivian Genaro Motti\textsuperscript{\textdagger}, and Sungsoo Ray Hong\textsuperscript{\textdagger}}
\affiliation{%
  \institution{\textsuperscript{\textdagger} George Mason University, Fairfax, VA \country{USA}\\ \textsuperscript{\S}Temple University, Philadelphia, PA \country{USA}\\ \textsuperscript{\textdaggerdbl} The Ohio State University, Columbus, OH \country{USA}}
  }

\renewcommand{\shortauthors}{Ara, et al.}

\begin{abstract}
Successful job search results from job seekers' well-shaped social communication.
While well-known differences in communication exist between people with autism and neurotypicals, little is known about how people with autism collaborate with their social surroundings to strive in the job market.
To better understand the practices and challenges of collaborative job seeking for people with autism, we interviewed 20 participants including applicants with autism, their social surroundings, and career experts.
Through the interviews, we identified social challenges that people with autism face during their job seeking; the social support they leverage to be successful; and the technological limitations that hinder their collaboration.
We designed four probes that represent major collaborative features found from the interviews--executive planning, communication, stage-wise preparation, and neurodivergent community formation--and discussed their potential usefulness and impact through three focus groups.
We provide implications regarding how our findings can enhance collaborative job seeking experiences for people with autism through new designs.

\end{abstract}





\maketitle
\section{Introduction}

When people are engaging in job seeking, their outcomes are often highly affected by their communication~\cite{ZIKIC2009117, Nikolaou_2014}.
To strive in the job market, people usually build a network, ask for help from their close social surroundings to prepare their applications, and communicate with employers to understand the context of recruitment~\cite{Wang2022-cz, Ferreira2023-se, MARTIN2021101741, Westbrook,hci}.
While leveraging one's social capital through well-planned networking and communication is crucial in job seeking~\cite{latham, WEBB19951113, articleohl}, people with autism may become easily overwhelmed in achieving this goal~\cite{neuro2, neuro3, davies2023}.
This occurs due to several reasons, but one predominant factor comes from the differences between the ways people with autism and neurotypicals (NTs) communicate~\cite{doi:10.1073/pnas.0910249107,koldewyn_weigelt_kanwisher_jiang_2012}.
The impact of such differences has been widely studied, for example, how people with autism experience difficulties in understanding social cues from the NTs~\cite{doi:10.1073/pnas.0910249107,koldewyn_weigelt_kanwisher_jiang_2012}, initiating conversation with strangers~\cite{nh} or interpreting nonverbal communication~\cite{doi:10.1073/pnas.0910249107}.
Such communication differences lead to social anxiety, resulting in people with autism encountering greater challenges in their job seeking process when compared to NTs~\cite{collab1}.

While prior studies have found the necessity of improving assistive technology or building an inclusive environment and workplace culture from the employers' perspectives~\cite{neuro1, emerg, inproceedingsweli}, there has been little focus on deeply understanding how people with autism collaborate with others to manage and coordinate their efforts in the job seeking process.
Exploring collaborative job seeking for people with autism introduces novel perspectives for Human-Computer Interaction research domains. 
Specifically, while a large body of research in Computer-Supported Cooperative Work (CSCW) focusing on learning how groups work in varying contexts of collaboration~\cite{hong2019design,zinatdata}, the ways people with autism coordinate their task dependencies~\cite{miranda1993impact, malone1994interdisciplinary}, define division of work~\cite{morris2007searchtogether}, build group awareness of activities and plan~\cite{heer2007design, hong2018collaborative}, and sharing skills, knowledge, and experience~\cite{jameson2007recommendation} under collaborative job seeking is not well understood.
Further, from the perspective of assistive computing, understanding the status quo of collaborative job seeking for people with autism provides new avenues of research for designing technologies that mitigate common challenges identified in former studies, including executive planning~\cite{LUNA2007474, liss_fein_allen, Van}, emotional regulation~\cite{samson2012emotion, MAZEFSKY2013679, Samson_Hardan, inproceedingsweli}, and communication differences~\cite{rubin2004challenges, biklen1990communication, picard2009future}.

\RTwo{make novelty explicit}
\TwoAC{Specify why the findings are not generic}
\TwoA{In this work, we extend our understanding of how the current CSCW-driven design, arguably built from the NT's perspectives imposes specific challenges for people with autism regarding collaboration in job seeking situations. 
Moreover, from the perspective of assistive computing, we aim to understand how different styles of communication between people with autism and NTs can introduce challenges for collaboration processes in job seeking situations.
Combining the two perspectives, we contribute to HCI by exploring what specific new technological designs can practically help people with autism collaborate with their surroundings leading to better outcomes in job seeking endeavors}.

To do so, we conducted semi-structured interviews with 20 participants (Study 1) \TwoX{to understand how and why people with autism collaborate with their social surroundings and when technologies cannot fully support their collaboration}.
To understand multi-faceted perspectives of collaborative job seeking practices, we identified three main stakeholders: (1) the individuals with autism who had/have experience in collaborating with their social surroundings (\textbf{Job seekers}), (2) their close social surroundings, such as their family members and friends who support the people with autism in their job seeking processes (\textbf{Surroundings}), and (3) career experts who have years of expertise assisting people with autism throughout their job search journey (\textbf{Experts}).
\TwoB{As a result, we identified 8 internal and external \textit{challenges} that people with autism encounter during their job seeking process and 8 corresponding \textit{strategies} indicating how the three stakeholders collaborate to address the challenges identified}.
\TwoC{We also identified the design space made with the 6 thematic \textit{needs} drawn from the 8 collaborative strategies, explaining when and how the current technology can not fully support their needs in collaboration}.

\RThree{Connecting Study 1, design space, and Study 2}
\TwoD{Based on the needs we found in Study 1, we selected four directions that require more dedicated system-level research and built prototypes as design probes.
The probes created represent (1) socially supported executive planning, (2) socially supported communication, (3) socially supported stage-wise job preparation, and (4) neurodivergent (ND) job networking}.
To understand the potential values of the directions, we conducted three rounds of focus groups, one with each stakeholder (Study 2).
\TwoE{The qualitative analysis reveals how the four directions can benefit future collaborative job seeking in practice for people with autism while also showing further considerations when applying the directions}.
\Three{Finally, we discuss how Study 1 and Study 2 provide design insights for current and future technology to facilitate collaborative job seeking. 
We then discuss opinions that may not directly inform the directions of design but are noteworthy.
The topics include the issues of social camouflaging, disability dongles~\cite{dongle}, ableism~\cite{ableism}, and reasons why tools built for neurotypical profiles may be insufficient to handle the ``socio-technical gap''~\cite{ackerman2000intellectual}}.

This work offers the following contributions.
First, Study 1 explains the reasons, strategies, and challenges that people with autism undergo when collaborating with their social surroundings in seeking jobs.
Second, we suggest the design space of collaborative job seeking for people with autism. 
Through four design probes and three focus groups, Study 2 demonstrates how designers can instantiate their ideas using prototypes perceived as useful. 
\TwoAC{In discussion, discuss the context of design. Don't reiterate related work.}
\Three{Finally, we provide implications for design that can inform researchers and practitioners in shaping the design of collaborative job seeking for people with autism in the future.}

\section{Related Work}

People with autism are eager to secure employment and contribute to society, however many individuals find themselves overlooked or underestimated in the job market.
Prior research shows that an inclusive work environment that considers employees with disabilities, i.e., hires people with autism is 30\% more productive than those who do not hire them ~\cite{Neuroarticle, dataautism}.
People with autism are highly creative, pay close attention to critical details others often miss, can think outside the box, and are more likely to report and address problems that could otherwise be unaddressed~\cite{cope2021strengths,nicholls2023autistic}.

To improve employment prospects for people with autism, several studies focused on understanding how the work environment can be transformed in a neurodivergent-friendly way.
For instance, \citet{muller_social_2008} found possible directions to facilitate communication between people with autism and neurotypicals in the workplace, such as finding communication topics centered around shared interests, preparing well-structured or scripted social events, providing instructions on how to interpret and utilize social cues, among others.
Aligning with this perspective, \citet{10.1371/journal.pone.0147040} observed that an employment environment tailored specifically for people with autism can result in higher levels of self-efficacy than non-autism-supportive work environments.
\citet{Ezerins_Simon_Vogus_Gabriel_Calderwood_Rosen_2023} highlighted the barriers and potential enablers that individuals on the autism spectrum encounter during varying stages of employment and steps that can be taken by organizations to mitigate it in pre-employment and post-employment stages.

\citet{Kim_2022} identified the need for increased structural support to foster collaboration among all stakeholders, rather than placing the entire responsibility on job coaches.
\citet{annabi_locke_2019} synthesized past literature into a theoretical framework to study IT workplace readiness and also revealed that barriers and opportunities in the IT workplace are influenced by individual differences, individual coping methods, autism employment programs, and neurotypical attitudes towards individuals with autism and knowledge about autism.
\citet{doi:10.1080/09638288.2018.1527952} conducted a comprehensive review of 829 articles from January 1987 to March 2018 to identify workplace accommodations that  support  employment  for adults with autism. The study results were categorized into five distinct environmental domains: (1) natural environment; (2) products and technology; (3) support and relationships; (4) attitudes; and (5) services, systems, and policies.

Another side of the research focused on developing interactive systems to support underrepresented populations who face disadvantages due to factors such as race, social class, gender, and ability in the job market.
\citet{10.1145/3476065} proposed a framework to improve the design of current employment support tools for marginalized job seekers.
Using these design requirements as a foundation, they developed a tool that helps job seekers to recognize their skill sets and find job openings that match with their skills~\cite{dreamgig}.
Further, Dillahunt and Hsiao~\cite{10.1145/3313831.3376717} conducted a field experiment on two tools. One named \textit{Review-Me} allowed job seekers to register their resumes and receive feedback from volunteers who agreed to serve as backup reviewers. Another named \textit{Interview4} provided a free online video tool that allows job candidates to refine their interview skills~\cite{10.1145/2901790.2901865, 10.1145/3313831.3376717, Interview4}. 
While the former approaches support underrepresented populations in general, HCI communities started to seek targeted approaches i.e., specifically built for people with autism.
For instance, \citet{10.1145/3505560} developed an adaptive virtual reality-based job interview training platform, \textit{Career Interview Readiness in VR}. 
\citet{10.1145/3472163.3472270} developed a job-matching system to match people with autism with employers after estimating how both would rank each other.
Aside from academic studies, various tools and resources started to support autism hiring ``in the wild'', including \textit{Mentra}~\cite{WinNT3}, \textit{Mindshift}~\cite{mindshift}, \textit{The Precisionist}~\cite{WinNT2}, \textit{Neurodiversity Network}~\cite{WinNT}, and \textit{SourceAbled}~\cite{sourceabled}.

While we found comprehensive research on (1) the ways employers can transform their workplaces in a more neurodivergent-friendly fashion, and (2) the system-level approaches that provide varying job-related support for people with autism, relatively few approaches have been found aiming to understand how people with autism are leveraging their social capital when seeking for a job. 
\TwoA{On the other hand, there have been several studies on understanding the relationship between social activeness and job-seeking outcomes for neurotypicals.
For instance, socially active seekers may have an increased chance of receiving offers by leaving stronger impressions on employers~\cite{Liu2014-xs}.
Also, job seekers who have support from their social surroundings can drive themselves to sustain their effort over an extended period~\cite{paul2009unemployment, schwarzer2007functional} or remain focused despite potential setbacks~\cite{Wang2022-cz} when compared with those who work alone.
\citet{dillahunt2018designing} performed a comparative validation study using 10 job search tools and found that the tools providing social support received most of the votes, demonstrating the practicality of motivating social activeness through new technology.
Exploring this body of literature, findings suggest that active communication and collaboration play a substantial role in improving the job outcomes of people with autism~\cite{10.1145/3173574.3173622, Bandura_2001, paul2009unemployment, schwarzer2007functional}}.
In this context, we are particularly motivated to understand more deeply  the current practices and challenges in the social collaboration between people with autism and their support networks.

\section{Study 1: Method}

Study 1 aims to understand when, why, and how people with autism collaborate in seeking jobs and when the current technology fails to support their specific collaboration needs fully.
To achieve this, we defined our initial directions of inquiry as follows:
(1) \textbf{challenges in job seeking}: what major challenges that people with autism are experiencing motivate them to seek collaboration;
(2) \textbf{collaboration practices}: what are the common advantages and coping strategies to deal with the challenges,
(3) \textbf{technological limitations}: what aspects of the current technology hinder the collaborative job seeking for people with autism, and
(4) \textbf{a wish list and desired features}: what desired features should be prioritized for motivating collaborative job seeking for people with autism.
We conducted open-ended semi-structured interviews as a method that allows us to personalize the questions based on the participants' experiences entailed in collaborative tasks of job seeking.

\subsection{Participants}
To capture multi-faceted aspects of collaborative job seeking for people with autism, we first identified the (1) \textbf{job seekers with autism} as a main stakeholder of collaborative job seeking. Next, we defined two other major stakeholder types who worked with the former group.
They were (2) \textbf{social surroundings}, people including family members, friends, or close colleagues who have worked with persons with autism and helped them to get a job, and (3) \textbf{experts}, including job coaches who are professionals in supporting persons with autism to get a job.
\OneA{To define the stakeholders' types, we analyzed related work, specifically literature that studied the relations between job seekers' social activeness and their employment~\cite{Liu2014-xs, paul2009unemployment, schwarzer2007functional, Wang2022-cz} to identify what types of people collaborate on seeking jobs}.

\OneA{Since there are no past statistical descriptors or demographics available for the types of stakeholders we identified, for recruitment purposes, we used convenience and snowball sampling strategies~\cite{creswell2016}}. 
In doing so, we contacted acquaintances who work for motivating employment for people with autism and their social connections.
They were directors in research-centered university student employment programs, officers in public and private organizations that provide employment support services for people with autism, hiring managers who are in charge of inclusive/disability hiring, and academic and industry researchers who have worked with the stakeholder types we identified.
In contacting them, we provided a flyer that explained the purpose of the study, the required experience to be a participant, compensation information of a \$30 gift card, and our contact point.  
We asked the acquaintances to share the flyer with potential participants.
For those who showed interest in our study, we asked them to fill out the survey to see if they have relevant experience on collaborative job seeking for people with autism.
As a result, we interviewed a total of 20 participants.
Among them, we had 8 job seekers with autism; 4 identified themselves as females and the rest of 4 identified themselves as males; ages ranged from 24 to 44 years.
Next, we had 6 social surroundings in our interviews. All of them identified as female, with ages between 19 to 67. 
Finally, we had 6 experts who were affiliated with 6 different local educational institutions, companies, or non-profit organizations.
Two of them identified as males while the other 4 identified as females.
Their age ranged from 26 to 62 years. 
Table 1 summarizes the details of our participants.

\renewcommand{\arraystretch}{1.2}%
\begin{table*}
\small
  \centering
  \begin{tabularx}{\textwidth}{p{0.05\textwidth} p{0.1\textwidth} p{0.15\textwidth} X}
    \toprule
        \textbf{PID}
        & \textbf{Types}
        & \textbf{Education}
        & \textbf{Profile} \OneX{(years of professional experience for Experts)}\\
    \midrule
    \OneX{J1} & Job seeker & Master's degree & A Ph.D. candidate who is searching for a job\\
    \OneX{J2} & Job seeker & Ph.D. & A researcher who is searching for jobs in the field of computing \\
    \OneX{J3} & Job seeker & Master's degree & A licensed social worker who was in the job market\\
    \OneX{J4} & Job seeker & Bachelor's degree & A recently graduated job seeker who is looking for his first job in IT field\\
    \OneX{J5} & Job seeker  & Master's degree & A behavior therapist who sought therapist jobs a few years ago\\
    \OneX{J6} & Job seeker & Bachelor's degree & A job seeker looking for systems engineer or project manager positions \\
    \OneX{J7} & Job seeker & Some College & A job seeker looking for high skill-low self management opportunities\\
    \OneX{J8} & Job seeker & Bachelor's degree & A technical program manager who sought jobs in the past\\
    \midrule
    \OneX{S1} & Surrounding & Some College & A college student who helped her cousin in preparing for jobs\\
    \OneX{S2} & Surrounding & Ph.D. & A university professor who is helping students on ND spectrum for jobs\\
    \OneX{S3} & Surrounding & Master's degree & A business owner who helped her family member to find a job\\
    \OneX{S4} & Surrounding & Master's degree & A mother who helped her son in preparing for jobs \\
    \OneX{S5} & Surrounding & Ph.D. & A mother (professor in learning technology) who helped her son in preparing for jobs\\
    \OneX{S6} & Surrounding & M.D. & An adjunct faculty who helped her husband in preparing for jobs  \\
    \midrule
    \OneX{E1} & Expert & Master's degree & An assistant director of student professional development at a university \OneX{(1+ years)} \\
    \OneX{E2} & Expert & Ph.D. & A director of center for autism at a private organization \OneX{(6+ years)}\\
    \OneX{E3} & Expert & Bachelor's degree & A director of autism career development at a nonprofit organization \OneX{(35+ years)} \\
    \OneX{E4} & Expert & Bachelor's degree & An occupational therapist and career advisor at a private organization  \OneX{(20+ years)}\\
    \OneX{E5} & Expert & Bachelor's degree & An employment specialist for people with disability at a service center \OneX{(7+ years)} \\
    \OneX{E6} & Expert & Bachelor's degree & An employment specialist of center for autism at a private organization \OneX{(15+ years)}\\    
  \bottomrule
  \end{tabularx}
  \vspace{2mm}
  \caption{
    A list of participants, from the left: (1) Participant ID, (2) Stakeholder type, (3) Education, and (4) their Profile and \OneX{years of experience (for experts)}}
    \Description{A list of Study 1 participants that explains four types of information, their Participant ID, Stakeholder type, education, and their profile along with years of experience on providing services for people with autism (for experts)}
  \label{tab:table1}
\end{table*}

\subsection{Interview}

All interviews were conducted online via Zoom by the the first and the last authors.
\OneA{At the beginning of each interview, we provided a consent form and asked participants to provide their agreement if they would like to proceed with the study. 
We recorded the conversation upon the acceptance from participants.
To provide consistent interview environments between participants, we prepared a slide deck that had the questions organized in the following 4 categories:
(1) general questions about their job seeking-related process: their challenges, strategies, and memorable episodes/stories; (2) their current technology use: types of tools they use for searching jobs or communicating with their social surroundings, and notable challenges related to technology usage; (3) their collaboration practices: how do they get and provide help, when they feel challenges in collaboration; and (4) their new technology needs with high priority: how future designs should evolve to improve job seeking technology and collaborative job seeking tools.}
\OneX{The interviews lasted 59.89 minutes on average (SD=18.51) while the longest one lasted 108 minutes and the shortest one 40 minutes.}
\OneX{After the interview, we informed our participants that the study findings would be shared with them for their review before becoming public.}
Every interview was transcribed by English-proficient transcribers.

\subsection{Qualitative Coding Process}
\label{coding}

\OneA{In our qualitative analysis, we followed an iterative qualitative coding process~\cite{saldana2015coding, layder1998sociological}.
In the first stage, three of the coders separately coded text segments from each transcription through the three stages of pre-coding, jotting, and final coding~\cite{saldana2015coding}.
As we conducted more interviews, the coders started building analytic ``notes'' and ``memos'' to identify recurring patterns and connect related insights among different stakeholders.
The coders updated the notes as they proceeded with the interviews and discussed weekly to share insightful remarks about the commonalities and discrepancies among participants.
They also checked if data collection had reached a saturation point regarding novel findings related to collaborative job seeking for people with autism.
In doing so, the authors gauged the common findings across the insights shared at the last interview as well as the insights learned from all the interviews~\cite{saldana2015coding}.
It took roughly six months for the three coders to agree on the saturation of findings.
The first interview took place on March 19th, whereas the last one was carried out on August 14th, 2023.
Lastly, the coders started building the unified structure of the seed themes and sub-themes by reviewing all the codes that emerged, as well as the quotes, memos, and snippets extracted with relevant details. This whole process was iterative across three rounds of discussion.}

\section{Study 1: Results}
\label{res:Study 1}

\TwoB{
We present the results of Study 1 findings (see table~\ref{tab:S1results}) based on our inquiries and three types of stakeholders' perspectives. 
We introduce \textbf{three main themes}: (1) \textbf{Internal challenges and collaboration practices} in job seeking, (2) \textbf{External challenges and collaboration practices} in job seeking, (3) \textbf{Technology needs and design space}.
The first theme describes various challenges stemming from inherent characteristics that people with autism face in their career search and corresponding collaboration strategies implemented to cope with those.
The second theme outlines the challenges arising from external factors and societal practices, along with coping mechanisms adopted by our participants.
In the third theme, we discuss our participants' preferences regarding tools' features and synthesize their specific technology needs as a design space aimed to improve collaborative job search experiences for people with autism and their peers.
}
\TwoAC{not clear about S1 findings}

\begin{table*}[htbp]
\small
\centering
  \begin{tabularx}{\textwidth}{p{0.12\textwidth} p{0.27\textwidth} p{0.27\textwidth} X}
    \toprule
    \TwoX{\textbf{Challenges}} & 
    & \TwoX{\textbf{\hspace{0.15cm}Collaboration Practices}}
    & \TwoX{\textbf{\hspace{0.15cm}Technology Needs}} \\
    \midrule

        \begin{tabular}[x]{@{}l@{}}\TwoX{\textbf{Internal}}\\\TwoX{\textbf{Challenges}}\end{tabular} &
        \begin{tabular}[x]{p{4cm}l@{}}
            \TwoX{\textbf{\textit{Executive functioning}}}\\
            \TwoX{Handling multitasking, task prioritization, memorizing a long list of information/conversations from job calls, and handling alarm fatigue}\end{tabular} &
        \begin{tabular}[x]{p{4cm}l@{}}
            \TwoX{\textbf{\textit{Simplifying complex tasks}}}\\
            \TwoX{Helping with the decomposition of complex tasks, step-by-step timeline organization, event reminders, tracking, and physicalizing important information (e.g., note down)} \end{tabular} &
        \begin{tabular}[x]{p{4cm}l@{}}
            \TwoX{\textbf{\textit{Collaborative executive planning}}}\\
             \TwoX{Integrated job events in a centralized way, smart reminder, design that can offload cognitive burden in multitasking, converting job posts to text-to-speech} \end{tabular} \\
    &
        \begin{tabular}[x]{p{4cm}l@{}}
            \TwoX{\textbf{\textit{Communication style difference}}}\\
            \TwoX{Reading context in conversation (as they are ``literal thinkers''~\cite{milton2012ontological,crompton2020autistic} ), idea abstraction, and interpreting subtleties/euphemisms from job calls} \end{tabular} &
        \begin{tabular}[x]{p{4cm}l@{}}
            \TwoX{\textbf{\textit{Translating NT languages}}}\\
            \TwoX{Helping with email reading \& writing, preparing interview scripts, and occasionally reaching out to employers for questions in advance} \end{tabular} &
        \begin{tabular}[x]{p{4cm}l@{}}
            \TwoX{\textbf{\textit{Collaborative communication}}}\\
            \TwoX{Groupware for collaborative reading and editing email, NT language translation feature for people with autism} \end{tabular} \\
    &
        \begin{tabular}[x]{p{4cm}l@{}}
            \textbf{\textit{\TwoX{Networking}}}\\
            \TwoX{Handling anxiety caused by cold conversation, maintaining a network, speaking in a large group, and fear of being perceived as ``rude''} \end{tabular} &
        \begin{tabular}[x]{p{4cm}l@{}}
            \textbf{\textit{\TwoX{Help in expanding networks}}}\\
            \TwoX{Training social and soft skills, connecting peer ND job-seekers or employees, and introducing virtual career fairs} \end{tabular} &
        \begin{tabular}[x]{p{4cm}l@{}}
            \textbf{\textit{\TwoX{Collaborative networking}}}\\
            \TwoX{Motivating the connecting with peers/mentors in ND spectrum, and establishing an intelligent soft skill learning and sharing resources}\end{tabular} \\
    &
        \begin{tabular}[x]{p{4cm}l@{}}
            \textbf{\textit{\TwoX{Emotional dysregulation}}}\\
            \TwoX{Handling rejections and imposter syndrome, balancing between required support and autonomy}\end{tabular} &
        \begin{tabular}[x]{p{4cm}l@{}}
            \textbf{\textit{\TwoX{Emotional support}}}\\
            \TwoX{Providing advice and sharing of experience for supporting emotional distress and recovery} \end{tabular} 
        \\
    \midrule
        \begin{tabular}[x]{@{}l@{}}\TwoX{\textbf{External}}\\\TwoX{\textbf{Challenges}}\end{tabular} &
        \begin{tabular}[x]{p{4cm}l@{}}
            \textbf{\textit{\TwoX{Shaping directions for career}}}\\
            \TwoX{Lack of specific career guidance, seekers' strong job-related preferences, and NT-targeted job descriptions limit the opportunities}\end{tabular} &
        \begin{tabular}[x]{p{4cm}l@{}}
            \textbf{\textit{\TwoX{Pre-assessment \& career coach}}}\\
            \TwoX{Providing a pre-assessment process for finding job adaptability and career coaching specific to the career direction.} \end{tabular} &
        \begin{tabular}[x]{p{4cm}l@{}}
            \textbf{\textit{\TwoX{Collaborative job preparation}}}\\
            \TwoX{Making group collaboration for preparing, proofreading, job materials, and facilitating mock interviews} \end{tabular} \\
    &
        \begin{tabular}[x]{p{4cm}l@{}}
            \textbf{\textit{\TwoX{Applying and interviewing}}}\\
            \TwoX{Adjusting resumes, flexibly improvising answers to unexpected interview questions} \end{tabular} &
        \begin{tabular}[x]{p{4cm}l@{}}
            \textbf{\textit{\TwoX{Helping on job preparation}}}\\
            \TwoX{Proofreading, eliciting resume feedback, and practicing mock interviews with their surroundings} \end{tabular} &
        \\
    &
        \begin{tabular}[x]{p{4cm}l@{}}
            \textbf{\textit{\TwoX{Using ``go-to'' job search tools}}}\\
            \TwoX{Reading ``vague'' job descriptions, using incomplete or not autism-friendly searching criteria, and assessing whether the diversity statement is ``boilerplate''} \end{tabular} &
        \begin{tabular}[x]{p{4cm}l@{}}
            \textbf{\textit{\TwoX{Connecting hiring managers}}}\\
            \TwoX{Eliciting more context directly from hiring managers on behalf of job seekers} \end{tabular} &
        \begin{tabular}[x]{p{4cm}l@{}}
            \textbf{\textit{\TwoX{Rethinking ``search'' features}}}\\
            \TwoX{Adding ND-friendly criteria, such as team size, communication frequency, providing sensory factors, i.e., noise level, temperature, etc.} \end{tabular} \\
    &
        \begin{tabular}[x]{p{4cm}l@{}}
            \textbf{\textit{\TwoX{Fighting against social stigma}}}\\
            \TwoX{``Thin slice judgments''~\cite{sasson2017neurotypical} about autism poses people with autism with the dilemma of masking themselves} \end{tabular} &
        \begin{tabular}[x]{p{4cm}l@{}}
            \textbf{\textit{\TwoX{Masking vs. self advocacy}}}\\
            \TwoX{Providing advice to people with autism to help them decide between masking and self-advocating} \end{tabular} &
        \begin{tabular}[x]{p{4cm}l@{}}
            \textbf{\textit{\TwoX{Technology cannot fix everything}}}\\
            \TwoX{Social stigma cannot be handled in a single design or interface, but more designs to facilitate broader social awareness about people with autism are required} \end{tabular} \\
  \bottomrule
  \end{tabularx}
  \caption{\Two{Study 1 results at a glance: \textbf{Internal and external challenges} in seeking jobs for people with autism (left), their typical \textbf{collaboration strategies} (middle), and \textbf{unmet technology needs} that could improve their current practices of collaborative job seeking (right).}}
  \Description{A table that explains the study 1 results, from the left, typical internal and external challenges that people with autism experience while seeking jobs, their collaboration strategies with their social surroundings and experts, and unmet technology needs that can improve their practice of collaborative job seeking.}
  \label{tab:S1results}  
\end{table*}

\subsection{\TwoX{Internal Challenges and Practices}}
\TwoAC{generic challenges, not specific to autism}



\TwoX{
Internal challenges refer to the barriers that Job seekers face due to their disabilities in various phases of their career search.
While past literature has examined certain challenges associated with autistic characteristics in general~\cite{muller_social_2008,10.1371/journal.pone.0147040, Ezerins_Simon_Vogus_Gabriel_Calderwood_Rosen_2023, annabi_locke_2019}, our study findings highlight how these traits impact their social collaboration in the job seeking process, a perspective not previously explored in HCI and CSCW community.
Details of these challenges are described in the following four sub-themes.
}

\subsubsection{\textbf{\TwoX{Executive Functioning}}}
\paragraph{\textbf{\TwoX{Challenges: }}}
Our participants talked about executive functioning challenges~\cite{LUNA2007474, liss_fein_allen, Van} related to \textbf{managing multiple tasks}, \textbf{switching between tasks}, and effectively \textbf{organizing resources}, time-constrained problem-solving that are cognitively demanding for individuals with autism.
\TwoX{Organizing each piece of information into its respective field during the job application and preparation phases increases their cognitive load. As J2 remarked, \textit{``Writing documents was fine, it was organizing, getting them submitted and getting the stuff into the right fields which was hard''}}.
\TwoX{Juggling multiple tasks and switching at the same time can be taxing for Seekers, as J1 mentioned, \textit{``I found the interview process for Ph.D. programs to be very energy-intensive, it's very much like you have to be switched on 100\% of the time for like 3 days, and it's very exhausting.''}}.
\TwoX{Consolidating information from \textbf{scattered resources and prioritizing} the tasks is also difficult.}
\TwoX{Five participants mentioned challenges regarding maintaining the necessary level of attention, which can result in them overlooking important information.}
The constant need to be vigilant for upcoming schedules, tracking and timely responding introduce \textbf{alarm fatigue}-type issues.
For example, J2 shared how too many alarm notifications can divert someone from doing the actual task, \textit{``There can be so many alarms going off all time that the one that warns about what somebody should be doing right now, does not capture their attention. So false alarms are very problematic''}.
A large percentage of people with autism have co-occurring ADHD~\cite{MICAI2023105436}, therefore\textbf{ time blindness} or losing track of time can manifest as missing important events or forgetting to perform tasks while focusing intently on another activity. 

J2 believes the traditional evaluation process requires reevaluation, as it frequently depends on memorization and rapid thinking. 
He remarked, \textit{``I think the typical process is fairly problematic because it’s a lot of on-the-fly stuff in your head where you have to just memorize stuff''}.
An individual on the autism spectrum often possesses innovative ways of thinking and problem-solving.
A candidate's capacity to memorize and recall answers during an interview may not accurately reflect their talent or suitability for a position.
This creates additional pressure on them because of their reduced working memory combined with their attempt to maintain a positive image during the interviews (J4).

\textbf{\TwoX{Collaboration Practices: }}
Job seekers find it beneficial to have social help in handling complex tasks in different phases of their career search~\cite{annabi_locke_2019, Lashkari_2023,10.1145/3476065}.
The involvement of social surroundings and expert professionals reduces the task complexity offering support in these areas.
E1 advised to \textbf{simplify complex tasks by breaking} them down into manageable and \textbf{easy-to-understand steps} to help with executive planning.
A step-wise \textbf{detailed plan with a timeline} and guidance for each task is useful, as E1 said, \textit{``At the end of an appointment, I want my students to come with a list of what they are doing next, and preferably a timeline of it''}.
\TwoX{Helping job seekers \textbf{organize the work materials in their personal preferred} way increases task efficiency.
For instance, J5 stated,  \textit{``I use a paper planner that breaks down my day into 30-minute increments so I can easily see the amount of time I need to dedicate to each task. Everything is laid out visually for me''}.
Family members also assist by staying updated on upcoming schedules and \textbf{providing regular reminders}, as expressed by S5 in assisting her son, \textit{``I would remind him periodically what to do next, how to prepare''}.
}

\subsubsection{\textbf{\TwoX{Difference in communication style}}}
\paragraph{\textbf{Challenges:}}
People with autism tend to interpret language and words in a very concrete and straightforward manner~\cite{hobson2012autism}.
E2 who also identified himself on the autism spectrum, emphasizes this issue, \textit{``We are very \textbf{literal thinkers}. If you say one thing with the literal definition of your words, but mean something else we are very likely to get confused''}.
In written communication, such as emails or documents, their perspectives are sometimes \textbf{interpreted as blunt} or straightforward (J2).
J1 shared her difficulties in writing emails: \textit{``Communication through email is always difficult for me because I hate writing emails to begin with, and I am not great at introducing myself''}.
Five job seekers also struggle to determine the \textbf{level of detail} in writing emails, for example, J2 said: \textit{``Sometimes when I am going over stuff in so much detail, it will exhaust the people I am talking to''}.
\TwoX{Deciphering the job-related subtleties and figuring out the unspoken rules is difficult for people with autism because these nuances are not explicitly delivered in a manner favorable to their understanding.}
J5 figured out that many job advertisements are not explicit about the actual nature of the role and use \textbf{euphemisms}, \textit{``Unfortunately, no one had taught me the subtleties of job searching and they don't say what they mean. That is how I got tricked into being a salesperson. I saw the word "client manager", and thought I would be managing clients''}.

\TwoX{During interviews, participants face open-ended questions that can have countless potential responses or questions with very generic concepts (i.e., where do you see yourself in 5 years). 
Such questions can create difficulties for autistic job seekers in articulating their thoughts and \textbf{ideas abstraction}.}
As J6 mentioned: \textit{``There are some standard interview questions like, “Tell me about yourself” that I find incredibly challenging as it has almost infinite possibilities''}.
The most commonly reported problem by our participants was the issue of communication with  neurotypical individuals.
A neurotypical behavior within a conventional social context can be comprehended incorrectly by autistic individuals.
As a result, job seekers tend to \textbf{overlook the context}, as J1 faced in interviews: \textit{``A lot of times, I miss the mark. It kind of derails the process of actually making that connection and getting the information that both of us need. Especially in interviews, the interviewer is looking for something, and I don't know what they are looking for''}.

\TwoAC{generic challenges, not specific to autism}
\OneAC{discussion should be deepen}

\textbf{\TwoX{Collaboration Practices: }}
\TwoX{To bridge the communication gap and reconcile the distinct interpretation styles between neurotypical (NT) and neurodivergent (ND) individuals, ``translating NT language'' can be an effective method.
Helping seekers to understand the neurotypical language and learn more about the real context of the question. 
As E2 stated: \textit{``One of the first things I kind of coach them on is what do people mean when they ask questions and stuff? So I coach people that you have to translate what they are saying to what they actually mean. They say one thing and they mean something else''}.} 
Many of our seeker group participants emphasized the usefulness of ``\textbf{scripting}'' as one of their preparation strategies for interviews.
J1 expressed her need for scripts to approach structurally interviews and email communication to ensure she effectively conveys her thoughts without feeling like she is bothering others. 
\TwoX{J8 mentioned creating extensive scripts using her past experiences,\textit{``For every interview I often prepped like 6 to 10 pages of a script. I used my experiences as a product manager and also research to figure out what likely questions people are going to ask''}.}
\textbf{Reading and crafting these scripts and emails} collaboratively with the help of social support is important because selecting appropriate wording and ensuring the right length of the content assist them in navigating communication more effectively.
An additional effective method for successful communication involves offering clear guidance for interview preparation, i.e., \textbf{sharing questions beforehand}, \textbf{organizing post-interview follow-up sessions} for job seekers to refine their responses, or \textbf{allowing additional time} during the interview.
As J2 said, \textit{``I don't like to think of all the things during the meeting. Getting the questions in advance can be very helpful. Or having a follow-up in two hours, I will have 10 other ideas and useful information I could give you''}.

\subsubsection{\textbf{\TwoX{Fear of Networking}}}
\paragraph{\textbf{Challenges:}}
Job seekers shared their experiences of having \textbf{social anxiety} and communication~\cite{rubin2004challenges, biklen1990communication, picard2009future} issues including \textbf{initiating conversation}, \textbf{mixing with other} people, and difficulties in understanding social cues and signals during their career search.
Some are uncertain about how to \textbf{connect with a network}, and may struggle with the nuances of self-promotion and marketing themselves to potential employers.
\TwoX{For instance, J5 said: \textit{``I know the vast majority of jobs are gained through networking. I don't know how to do that without coming across as pushy or just turning people off''}.}
Initiating communication with \textbf{``cold messages''} and articulating the message's content is also a challenging aspect.
\TwoX{Sometimes unexpected social situations such as \textbf{speaking in a large group} become a burden for them due to nervousness}.
For example, J1 felt stressed while attending a mandatory post-interview gathering, stating that: \textit{``Straight after all the interviews they go to a bar and say, "Okay, now socialize." It is supposed to be like a "de-stress" sort of thing, but socializing is really distressing for me''}.
Few participants also blamed the lack of emotional intelligence and social skills as blockers to their personal growth in fighting social awkwardness for which neurotypicals may \textbf{perceive them as "rude"}.
\TwoX{As a result, they struggle to cultivate important soft skills such as being a team player, assisting and providing support to other people when needed which are essentially required in any kind of workplace.}

\textbf{\TwoX{Collaboration Practices: }}
12 participants highlighted the importance of \textbf{forming connections with fellow neurodivergent individuals}, which can be effectively facilitated through social support.
Expert group participant E3 is actively engaged in facilitating the connection between the ND community to find ND-peers, \textit{``I always want people to find other neurodivergent people who get them, with whom they can unmask. I always want to support people in finding others who understand them''}.
J6 talked about connecting with an entrepreneur who has a similar trait,\textit{``I felt a great connection with him because we have a similar profile. We both lack a bunch of social skills. We can talk about any topic and be on a similar level''}.
Peer connection is helpful for new job seekers as J4 mentioned, \textit{``It is better to make more connections with people who are neurodivergent candidates and who were able to get a job. To know what tips they have, or what employers to look for''}.
The expert group participants significantly contribute to expanding the seeker's external network by introducing them to different career events.
Online programs and virtual events tend to be more user-friendly and less intimidating for individuals with autism compared to in-person events. 
J3 mentioned, \textit{``I learned about a \textbf{virtual career fair} through my school. It sounded appealing because that's much less daunting than having to walk up to tables and try to talk to crowds of people''}.
To mitigate the communication difficulties between neurodivergent and neurotypical individuals, expert group participant E2 shared their strategies for providing \textbf{training soft skills} to the seeker group, \textit{``The other side of that coin is learning what to say and what not to say, how to interact with people in a way that we don't come across as rude or blunt or too direct or shocking''}.

\subsubsection{\textbf{\TwoX{Emotional Dysregulation}}}
\paragraph{\textbf{Challenges:}}
The impact of \textbf{rejection} can be intensified for people with autism due to heightened sensitivity to stimuli, leading to emotional distress~\cite{samson2012emotion, MAZEFSKY2013679, Samson_Hardan}.
The pursuit of employment often triggers a cascade of dreams and aspirations, however, for ND individuals navigating the job market in a neurotypical manner, the emotional toll can be profound which is inextricable from tasks. 
The true cost lies in the emotional burden, i.e., depression, hopelessness, \textbf{feelings of ineptitude}, and a l\textbf{ack of self-worth}.
For instance, J6 mentioned: \textit{``They wanted me to apply for 10 different jobs every week, which is expensive for my energy levels. If I get a second or third interview, I get excited and start creating fantasies about what my life is going to be like in 5 or 10 years. So being rejected 10 times is very frustrating and painful to apply for a new job''}.
\TwoX{S5 talked about her son's emotional struggles after deciding to discontinue his college education: \textit{``So, college was a struggle for him, we brought him home. I think it was a difficult time for him because he was frustrated and did not feel successful in the world''}.}
Despite investing considerable effort inadequate follow-up updates regarding applications lead to increased anxiety affecting their daily routine.
As J1 said: \textit{``It is like I am just spouting out into the world and not getting any feedback''}.

\TwoX{Another important observation is that individuals with autism \textbf{seek autonomy} to make decisions and crave independence in their actions. 
Achieving this can be challenging, and relying on the surrounding's assistance for every decision can lower their self-esteem.
S5 remarked on her son's feelings in this regard, \textit{``He wants to stand on his own and he does not want his mother telling him how he should act and what he should say''}.}
\TwoX{Constantly worrying about performance elevates stress due to a history of \textbf{social criticism} and negative comments.}
S5 mentioned, \textit{``\textbf{Imposter syndrome} or just thinking that we are not doing a good enough job, those anxieties are even more pronounced for these individuals because they have had so much negative feedback their whole life''}.

\textbf{\TwoX{Collaboration Practices: }}
\TwoX{``Emotional support from surroundings'' during hard times helps job seekers to recover from their distress.}
For example, surrounding group participant S5 shared her experience with her son: \textit{``Unfortunately, he did not get picked for the cohort and it was very stressful for him and us too. But we did our best and I tried to manage his expectations, saying `It's okay if it does not work, it is a good experience to have and I am really glad you did that'''}.
Expert professionals consider \textbf{motivating} and \textbf{encouraging} individuals on the spectrum as an essential part of their career advising. 
For instance, E1 mentioned: \textit{``When I coach students I act as a cheerleader and make sure they feel supported through every step of the process. A big part of my job is bolstering confidence and reducing anxiety during a stressful and uncertain time in their life''}.
E2 emphasizes the importance of \textbf{investing time in self-reflection} to identify their strengths, \textit{``We have some issues with emotional regulation. We get stressed and nervous about things. It has been my general experience that the most successful autistic adults are those people who have taken some time and effort to figure out what their strengths are, and what they have to leverage that makes them a real asset to the workplace''}.

\subsection{\textbf{\TwoX{External Challenges and Practices}}}

\TwoX{External challenges are not directly connected with the autism characteristics of an individual and can be influenced by some other factors such as job seekers' skill lacks, social culture or stigma, and neurotypical perspectives. Details of these challenges are described in the following four sub-themes.}

\subsubsection{\textbf{\TwoX{Shaping directions for career}}}

\paragraph{\textbf{Challenges :}}
\textbf{Lack of guidance}, and \textbf{being unaware of the process} prevent job seekers from effectively leveraging the resources and connections in choosing a definitive career path.
\TwoX{ S6 from the surrounding group expressed: \textit{``I think the hardest thing is, everything out there is set up as if you already know what you want to do. Whether it's a doctor, a lawyer, a pipe-fitter, or an electrician, I don't know what job will be easy for him''}.
Some participants felt \textbf{overwhelmed when reviewing the job descriptions} and felt low confidence in their skills}.
For example, J6 said: \textit{``I think my bigger challenge is that I don't have the right degree and work history''}.
Meeting individual criteria and\textbf{ specific job preferences} limits the pool of available options for them and \TwoX{makes it difficult to find the best-matched results.}
\TwoX{For instance, wanting specific accommodations in desired work arrangements, such as remote work, reluctance to relocate, or aversion to smaller companies, make it more complex to cater to everyone's specific requirements.
For example, J4 said: \textit{``If it is reachable by public transportation, I could move there. But right now, I prefer to be 100\% remote''}.}

\textbf{\TwoX{Collaboration Practices: }}
Conducting ``pre-assessments and coaching services'' at the beginning of a job search is important, serving as a foundational step to customize job search strategies and employment opportunities to align with the Seeker's needs and strengths.
S5 emphasized how a comprehensive assessment provides insights about an individual's profile, for finding \textbf{job adaptability}, \textit{``Neuropsychological assessment is so crucial because when they reach adulthood point, there is hardly any support out there for these kids. I would say if your child has more severe symptoms and attributes then you need that assessment as it could be shared to future employers''}.
E4 mentions her advising begins with a \textbf{functional assessment}, where they consider various aspects of the individual's life outside of work, such as their living situation and overall well-being.
Along with the same line of thought, E6 pointed out that the \textbf{initial intake process} of collecting information about the individual's background, experiences, and aspirations helps them to understand their challenges and needs.

\subsubsection{\textbf{\TwoX{Applications and Interviews Preparation}}}

\paragraph{\textbf{Challenges:}}
Job seekers expressed their difficulties at various stages of preparing application materials, including the \textbf{adjusting of resumes}, cover letters, and email drafts.
For example, J1 mentioned her challenges when it comes to structuring a resume, \textit{``I know that many people are using applicant tracking systems. I have been trying to structure my resume so it becomes easy to read, but I haven't been able to find any information on how to do that''}.
\TwoX{J2 believed that interviews are structured unfairly because job seekers on the spectrum have to adopt a neurotypical script to create a favorable impression}.
\textbf{Improvising an answer} to the unexpected questions in the middle of the interview is difficult for them.
People with autism often rely on prepared scripts to navigate interviews, and when confronted with questions out of their scripts, they face significant challenges in forming appropriate responses.
As J4 said, \textit{``Sometimes, they throw out a question unexpectedly that might not be possible for me to answer immediately''}.

\textbf{\TwoX{Collaboration Practices: }}
The most common and effective strategy we observed is making ``social surroundings involved in the preparation process'' throughout the job search journey.
For example, J1 shared, \textit{``I had my parents look over my resume, and they \textbf{gave me feedback} on it. My dad works in IT, so he understands the technical aspects of things. My mom is pretty good at people skills. She helped me figure out how to word things on the resume so that it sounds more confident, but not overconfident''}.
J6 mentioned leveraging a friend's support as a direct reference that helped his application bypass the computer-based filtering system and get called for the interview.
The experts are involved in monitoring the progress of job applications, \textbf{proofreading} and providing more frequent updates to ensure the seeker's needs are met and their applications are progressing smoothly.
As E1 said, \textit{``For the neurodivergent population I tend to have more frequent email communication and appointments to follow up on how the job applications or interviews have been going''}.
She also mentioned offering valuable resources and supportive materials, including resume templates, \textbf{practicing likely interview questions}, and career assessments to prepare the seekers for their job hunt.

\subsubsection{\textbf{\TwoX{Using ``Current'' Job Search Tools}}}

\paragraph{\textbf{Challenges:}}
Interestingly, the majority of our participants expressed frustration and mixed feelings while using widely known job-searching platforms such as, LinkedIn, Indeed, Monster.com~\cite{linkedin, indeed, mosnter, ziprecruiter} etc. 
According to their opinion, these tools are\textbf{ not ND-friendly} and the most disliked features include \textbf{vague search criteria, filtering options, matched outcomes, complex navigation, information overload, and social media-like attributes}. 
Here are some statements of their feelings regarding these tools:
J3: \textit{``I found Linkedin and those sorts of things exhausting. It's really hard to find things on them. You search for a term, and like 80 things that are relevant come up''}.
J4: \textit{``On LinkedIn It's very unclear. What a free level position requirements are like, It's even harder to find a good job from a company that is accepting of neurodivergent candidates like myself''}.
J5: \textit{``That LinkedIn has struck me like a professional version of Facebook. It seems to be more socially oriented and geared towards growing connections, and getting a position in that manner. I have tried Zip Recruiter and Monster.com. None of them are user-friendly or comprehensive''}.
E4: \textit{``I am very much discouraging people applying through Indeed because it is a black hole''}.
S4:\textit{ ``In Indeed, you really don't know what you are getting, it is vague''}.
E4 raised concerns about ONET by the Department of Labor resources~\cite{ONET}, that this system is designed for a broad audience and the suggestions may not be suitable for a person with autism.

\textbf{\TwoX{Collaboration Practices: }}
\TwoX{\textbf{Directly ``contacting hiring managers''} during the search phase can be an effective approach.
Relying solely on standard job search tools may offer a general overview of available positions, but it might not provide the complete picture of a workplace's dynamics.
Career professionals possess the ability to directly engage with companies, gaining insight into the actual working environment. 
For instance, E5 emphasized the effectiveness of in-person contact over online applications, stating, \textit{``I start by checking Indeed for available positions, then use Google Maps to gauge proximity. I also explore other businesses that might not be actively hiring but make direct contact with their managers, introducing my participant. Speaking directly to the manager has resulted in more success than simply submitting applications on Indeed''}.
By leveraging their direct communication with companies, career coaches and specialists can gather invaluable information about job opportunities, assess workplace cultures, and promote neurodiversity hiring initiatives.}

\subsubsection{\textbf{\TwoX{Fighting against Social Stigma}}}
\paragraph{\textbf{Challenges:}}
\TwoX{Neurotypical people misinterpret certain behaviors and generalize based on preconceived notions about individuals with autism, their abilities, or work performance, instead of recognizing the qualities of each person.
As autism is an invisible disability, individuals who do not exhibit outward signs may face skepticism or disbelief about the challenges they experience.
Neurotypical people are not as willing to interact with people with autism based on thin slice judgements~\cite{sasson2017neurotypical}. 
This can make it necessary to mask their autistic traits, despite the negative mental health impacts.
J8 had the same perspective as well, expressing that: \textit{``They would have these judgments about me that would affect my overall performance even though I did amazing on things like take-homes and like actual concrete work they wanted from me''}.}
This introduces the biggest dilemma faced by our participants revolves around the decision of whether to ``mask'' and be ``upfront'' about it.
As there are chances for both positive and negative consequences associated with it. 
For instance, J2 mentioned that paying bills and financial obligations may cause him to mask in job interviews, \textit{``Nowadays I disclose more than I don't. I might not do that if I really needed the job to pay bills''}.
S4 from the surrounding group shared her perception and uncertain feeling of disclosing about her son, \textit{``He is a perfect example of someone who is extremely intelligent and super-capable of working independently but will struggle with the interview process and job seeking on his own. So we put him on LinkedIn and disclosed his neurodiversity. That is very difficult because I am sure it stops a lot of recruiters from contacting him''}.
\TwoAC{generic challenges, not specific to autism}
\OneAC{discussion should be deepen}

\textbf{\TwoX{Collaboration Practices: }}
\TwoX{Offering advice regarding ``masking or self-advocacy'' can aid job seekers in identifying the most appropriate approach for their career paths.}
\TwoX{Experts group participant E2 focuses on two major things while providing advice: \textit{``the most important bit, is self-advocacy. How do you advocate for yourself in a way that the people at work understand and empathize with what you need and are more likely to give you accommodations''}.}
E2 emphasizes to use of descriptors rather than labels when advocating for oneself. 
He explained that simply saying \textit{``I have autism''} might not effectively convey the specific challenges or needs a person has, and it can lead to misconceptions.
Instead, using descriptors like \textit{``I have a neurological condition with sensory processing challenges''} or being explicit about the accommodation need can allow the individual to explain their situation in more concrete terms.
E4 shares the same opinion and states that if seekers can follow a subtle approach to disclosing their needs rather than being explicit about their neurodivergent status, it may help them feel more comfortable and potentially avoid bias or misconceptions.

\subsection{\TwoC{Technology Needs and Design Space}}
\label{sec:dsnspace}

\TwoX{The final theme delves into the future design needs expressed by the stakeholders. 
We consolidate these insights to identify unaddressed technological requirements that could improve future practices of collaborative job seeking for people with autism. 
We identified 6 design needs driven by 8 challenge types. 
Each challenge has its corresponding needs, except the two pairs of ``networking'' and ``emotional dysregulation'' (the two corresponding to ``collaborative networking'') and (2) ``shaping directions'' for career and ``using go-to job search tools'' (the two corresponding to ``collaborative job preparation''), see table~\ref{tab:S1results}. 
In elaborating on the design space, we focus on the four needs among the six because the four are more suitable for drawing new design ideas at an HCI research level.
The rest of the two needs have either an implication for the current designs (i.e., ``rethinking search feature'') or require a more sociocultural approach that is beyond the scope of design (i.e., ``technology cannot fix everything'').
Therefore, we will discuss the rest of the two in section~\ref {sec:implication}.}

\subsubsection{\textbf{\TwoX{Collaborative Executive Planning}}}
has emerged as the most prominent collaboration suggestion
in Study 1 as job seekers felt they were investing more energy into ``planning'' and ``juggling'' tasks compared to neurotypicals. 
We identified that their struggle can be reduced by collaborating with Job seekers and Experts. 
However, the scattered information across different email threads, services, and websites makes it challenging to grasp the whole situation. 
In resolving this gap, one widely discussed solution is to help the group by providing access to job-related information in one place.
As E1 emphasized of \textbf{\textit{``centralized''} information access:} \textit{``..If there is a list of actions where they can check mark what is done and what is still to do, if they have one place to keep track of all of the jobs they have applied like, Okay, I've submitted my resume here, here, cover letter for these places''}.

The second dominant suggestion was understanding how to \textbf{leverage AI to organize scattered information} and inform groups about upcoming tasks.
S5 mentioned: \textit{``If you had an AI capability that could help them process information more quickly, organize information and help them execute tasks like step-wise problem solving''.}
To assist job seekers in processing information and managing their tasks, participants felt that AI can \textbf{extract crucial details from job postings}, including application deadlines, requirements, and company information, and then generate step-by-step task lists for each job application.

The last notable feature discussed was the need to help seekers handle time-sensitive tasks and deadlines using \textbf{smart reminders} as some job seekers can not regularly check their e-mail and miss deadlines.
Issues related to \textit{``Time blindness''}(J2) and \textit{``time management''}(E3) have been extensively brought up in Study 1. 
While notifying seekers about the upcoming deadline is important, designing such features can introduce alarm fatigue (J2).
Some individuals live precise to-the-minute schedules therefore the reminders have to be carefully calibrated to individual needs, i.e., using a distinguishable alarm or personalized sound notification from generic ringtones to capture their attention quickly. 
A feature to \textbf{convert text to speech}, especially when going through heavy text-based content on job portals is also considered useful (J2).

\OneAC{emphasize specific challenges spectrum perspective}



\subsubsection{\textbf{\TwoX{Collaborative Communication}}}

 is the second direction. 
``Lost in translation'' has been a topic discussed nearly as dominantly as executive planning. 
Our participants on the spectrum shared several cases of misinterpretation when they communicate with neurotypicals where they often interpreted as\textit{``rude''}, \textit{``blunt''}, \textit{``aggressive''}, \textit{``entitled''}, \textit{``arrogant''},  or \textit{``unfeeling monsters''}.
\TwoX{We discovered multiple methods employed by job seekers to seek assistance from both their social circles and experts, sometimes it can be everyday life advice. 
Social surroundings and career experts can serve as interpreters for translating NT language, \textbf{converting NT-friendly job advertisements} into formats more accessible for neurodivergent individuals. Introducing AI agents into this process will aid in learning these styles and autonomously suggest enhancements to achieve the desired format.
\textbf{Groupware} for \textbf{proofreading, editing and writing emails} can serve the purpose.}


\subsubsection{\textbf{\TwoX{Collaborative Networking and Mentoring}}}
 
 can be the next design direction as several participants discussed the importance of networking in seeking jobs with people who are beyond their reach. 
Not surprisingly, peers or mentors on the autism spectrum are the ones seekers were able to connect with comfortably, and those who \textbf{volunteered to assist the job seekers}, became incredibly valuable resources. 
\TwoX{Yet, it remains challenging for neurodivergent individuals to find such personals and connect with people whom they can trust and share experiences.
We believe that establishing a peer-neurodivergent network specifically dedicated to the neurodivergent community would greatly facilitate effortless connections and socialization further than connecting only with Surroundings and Experts.
As J1 said,\textit{``So maybe I just click with other neurodivergent people''}.
Individuals on the spectrum who have successfully secured a job can sign up as mentors and volunteer their assistance to those seeking jobs through this network.
Each mentor can be paired up with one or more seekers to
engage in direct conversations with them, guide them to \textbf{acquire necessary soft skills}, and provide or \textbf{share useful resources}.}

\subsubsection{\textbf{\TwoX{Collaborative Job Preparation}}}
includes reviewing job materials and preparing for the interview collaboratively which can highly increase the chance of success. 
In particular, resume fine-tuning can be exhaustive and taxing for several job seekers especially when they have to interpret subtle nuances about the different job calls and match with the expertise that they can offer.
\TwoX{J7 highlighted the advantage of preserving previous versions of the resume while editing a new one, \textbf{facilitating easy access to past edits} if needed, \textit{``When writing a resume, you omit certain things because they're not necessary for that particular job. It would be great to delve deep into my past and have a comprehensive record of everything I have ever done''}.
A feature where the group can work together offering meaningful \textbf{improvement in adjusting small details} on a resume, will be useful.}

\TwoAC{not clear what specific challenge is supported}

\TwoX{Our participants also brought up several suggestions while discussing interview-related issues. 
The general perception of Seeker's side about the current interview preparation materials is text-heavy and static rather than multimodal and interactive.
One notable pattern found in Study 1 is the importance of the script.
Compiling a list of common and potential interview questions, scripting answers, and engaging in \textbf{thorough practice sessions} can reduce anxiety.
A system capable of generating \textbf{a wide array of expected and unexpected questions} would enable seekers to practice comprehensively.
While experts also acknowledge the importance of presenting using multi-modal and interactive, understanding the design that can effectively motivate Seekers' learning experience in interviews remains elusive, and further research is required.}

\section{Study 2: Method}

\OneA{Study 2} \OneX{aims to validate and comprehend the impact of the new design spaces identified in~\ref{sec:dsnspace}, which delves into technology needs for potential enhancements in collaborative job searches.}
\TwoD{We selected the 4 needs that require more dedicated research from there and synthesized features to build an example prototype with 4 design probes, which are presented in the first subsection.
Finally, we conducted three rounds of focus groups with the three different stakeholder types of Study 1.}

\OneAC{connection should be strengthened}
\RThree{challenging to draw connections between findings}

\subsection{\TwoD{Synthesizing into Design Prototype}}

\begin{figure*}
  \centering
  \includegraphics[width=0.96\textwidth]{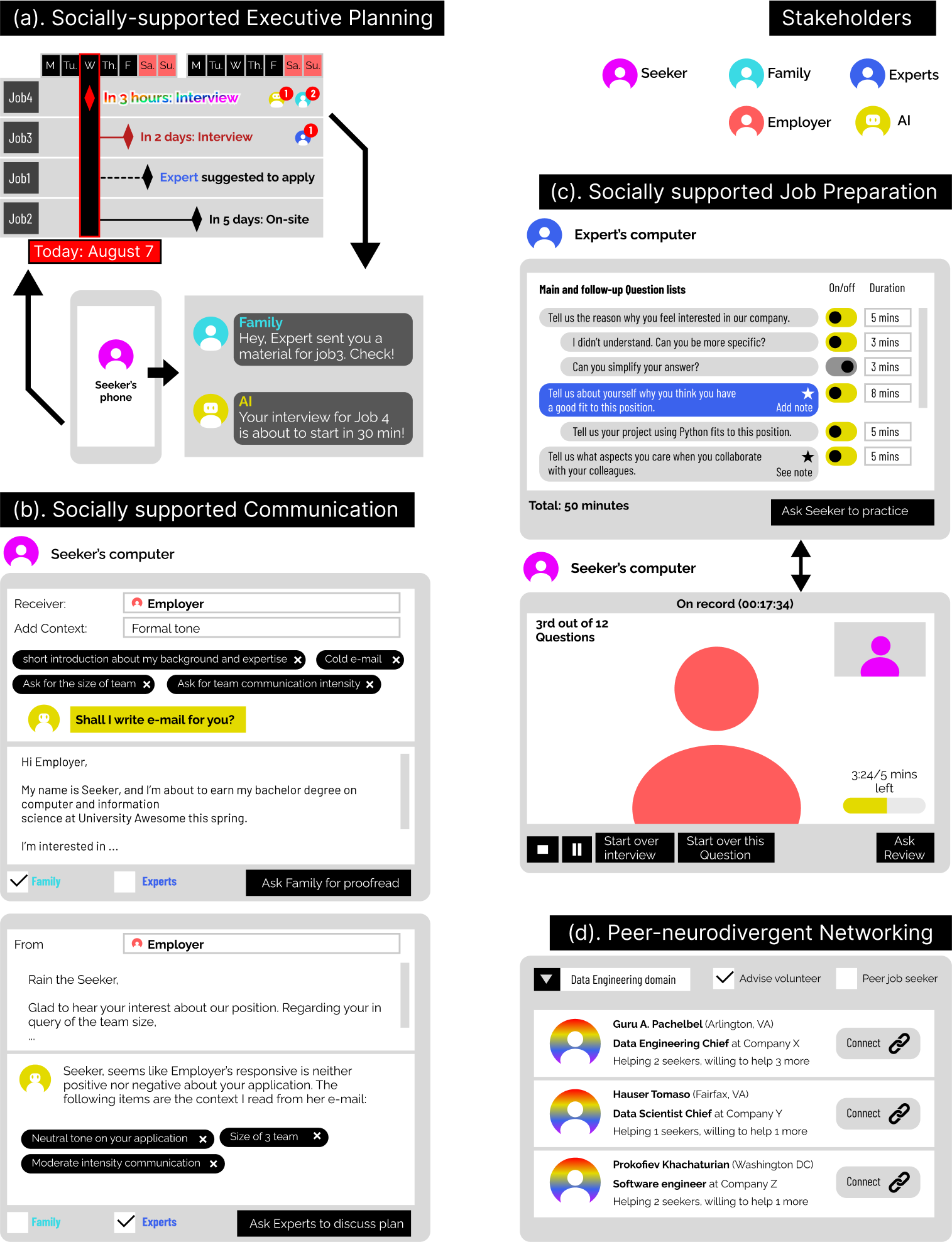}
  \caption{Four \OneX{design probes} from the design space of Collaborative Job-seeking for People with Autism.}
  \vspace{-0.2in}
  \label{fig:fig_prototypes}
  \Description{Four prototypes from the design space of Collaborative Job-seeking for People with Autism.}
\end{figure*}

\TwoX{We built a low-fidelity prototype with a set of {design probes} as an example of technology needs derived from the design space~\ref{sec:dsnspace}.
The design probes} include the following directions: (1) Socially supported executive planning, (2) socially supported communication, (3) socially supported job preparation, and (4) peer-neurodivergent networking.
In developing the probes, we defined human users, as well as AI agents as follows:  ``Seeker'', ``Family'', ``Expert'', and ``Employer'', and ``AI'' respectively (see Fig.~\ref{fig:fig_prototypes}, Stakeholders).


\textbf{Socially supported executive planning} feature is shown in Fig. \ref{fig:fig_prototypes} (a) as one of the design probes we built. 
\TwoX{The top screen shows the centralized timeline view of job lists that Seeker is working on.
The view is shared with other human users.
The design will prioritize highlighting the closest job deadline.}
Human users can send the message to Seeker and the messages can also be sent to Seeker's mobile (see bottom screen). 
Family and Expert can also set up job-related notifications on Seeker's devices leveraging AI. 


\textbf{Socially supported communication} \TwoX{ design supports collaborating on emails to mitigate the communication differences between people with autism and employers.} 
Fig. \ref{fig:fig_prototypes} (b), top shows the writing help. Seeker can add multiple word boxes as the context of the message. 
Upon adding a word box, Seeker can use AI to generate e-mail messages and revise them. 
Seeker can decide if she wants Family or Expert to proofread. 
Fig. \ref{fig:fig_prototypes} (b), bottom explains reading help. AI can interpret the core context and meaning of the message from the Employer.
Seeker can decide to share the received message with Family and Expert in case she wants to establish the plan together.


\textbf{Socially supported job preparation} \TwoX{feature facilitates collaborative resume fine-tuning and mock interviewing. 
Based on the job descriptions, AI will generate some topics about what aspects Seeker should focus on their resume to apply for that job.
Family will also engage in the process and can leave comments or suggest edits and share with Seeker.}
Fig. \ref{fig:fig_prototypes} (c), top and bottom screens show the interview prep. feature.
On the top, Expert or Family can generate a list of ``core'' questions (1st depth questions, such as ``Tell us the reason why you feel interested in our company''). For each core question, they can add ``follow-up'' questions (2nd depth questions, such as ``I didn't understand (your answer), Can you be more specific? '').
They can set the proper duration, add some useful notes for answering each question, and decide which questions they will include in the final set. 
Upon finishing listing up questions, they can send it to Seeker. 
The bottom screen shows the video conference mock-up interview from Seeker. 
The AI will randomly choose the question and ask. 
Seeker can answer it. 
While doing so, \TwoX{she can record}, check the time, pause, continue, and start over the question, or the whole interview session.
Once the mock-up is done, she can decide to share the interview with Family and Expert for review. 


\textbf{Peer-neurodivergent networking} \TwoX{feature is shown in Fig. \ref{fig:fig_prototypes} (d) where the interface can help Seekers find a peer neurodivergent job-seeker or a job mentoring advisor and an option to connect with them.
Seekers can also choose specific domains to shortlist their preferences, the system will retrieve the contacts based on the selected expertise.
Mentors can provide details of how many job seekers they are willing to advise.}

\subsection{Focus group Interview Process}

\OneA{Focus group interview was conducted in an open-ended fashion and the goal is to understand the following:}  (1) the priority and importance of the four design directions selected from Study1 and (2) presenting the four probes to the three stakeholder groups to know if instantiation of the directions can help the future job-seeking process of people with autism.
To achieve this, we conducted three rounds of focus group interviews with our three different stakeholder types in Study1: Job seekers, Surroundings, and Experts. 
\OneX{We reached out to the same participants using the same approach as in Study1 and consequently, from the interested candidates we recruited two Job seekers (J1, J4), three Surroundings (S2, S4, S5), and three Experts (E1, E2, E3).}
\OneAC{provide details about methods and participants}

For each design idea, we allocated 30 minutes to share and discuss the direction derived from design space and then introduced the probe as one of the examples that could handle the situation in the future. 
The expected time for each session was 2 hours. 
After that, we asked participants to discuss two aspects: (1) sharing their perspectives on the direction of the design space; including importance, priority, and postnatal usefulness as a direction for designing future groupware for collaborative job-seeking for people with autism, and (2) sharing their thoughts on the probe and specific suggestions that what can be done in each design to build the practical product.
Participants were compensated with \$30 of the gift card for contributing to the study.
\OneAC{data analysis of Study 2}
\OneX{Each focus group session was recorded and transcribed by professional transcribers. 
Three authors used the same method as Study1~\ref{coding} to do the analysis and collected feedback on two discussion sessions.}

\section{\TwoE{Study 2: Results}}

\subsection{\TwoX{Socially supported executive planning}}

\TwoX{Our participants found the centralized timeline view effective, in offering structured guidance and assistance for managing tasks in a timely fashion addressing the issues of ``time-blindness'' and ``multitasking''.
For example, S5 said, \textit{``For ASD job seekers, that would be very useful seeing four jobs, four applications going at the same time. I think that solves multitasking issues and as well as the time blindness''}.
J4 commented that the design is straightforward and accessible for individuals with autism who may feel uneasy when dealing with intricate systems such as enterprise software.
The job seekers liked the audio notification feature allows them to personalize it with a unique ringtone, aiding in easily distinguishing job-relevant notifications from the other reminders.
However, our participants raised few concerns about the executive planning design, i.e., ensuring proper training, the chance of getting overwhelmed by too many notifications, and potential anxiety from any unknown or new feature.}
For instance, E2 said, \textit{``There is going to be a learning curve and the challenge would be how to help people get over that. I have talked to many individuals who are desperate to find a way to overcome their executive function issues. They are sick of missing appointments, forgetting things, or multitasking. If people could be educated on how to properly use this, for some people, this could be a game-changer''}.
\TwoX{S5 expressed doubts about finding the optimal settings for notifications for ND job seekers, \textit{``It can tip over into becoming more anxious because both the family member and the career specialists are prompting him, So I don't know what will be the sweet spot of getting too many prompts''}.}

\TwoX{J1 and E3 suggested the option to integrate existing professional accounts (i.e., LinkedIn, Indeed) to retrieve profile information and populate job advertisements, and pulling events from calendars (i.e., Outlook, Gmail) will minimize the need for redundant data entry.}
S4 highlights the need for a confirmation mechanism that ensures acknowledgment when someone receives and views a prompt, \textit{ ``Maybe there needs to be a response like, "Got it," "Read it," "Saw it," so the person knows that it was read and received''}.
\TwoX{Supporting multi-modality will be beneficial as J1 said, \textit{``If it would send notifications to my smartwatch, that would be awesome because I am always wearing that I don't have my phone with me''}.}
\TwoAC{evaluation seems generic}

\subsection{\TwoX{Socially supported communication}}
\TwoX{The majority of the participants felt combining both AI and human perspectives could be comprehensible to mitigate 
misunderstanding between NTs and NDs.
S4 said, \textit{``I like it because it is using the strengths of both. Being able to pull out pieces for clarity that are in the email''}.
Involving AI in analyzing and generating the appropriate tone for the emails based on the user's inputs, addresses the challenge of initiating cold conversation.
E2 who is also on the autism spectrum, highlighted the design's potential as a valuable resource for enhancing email writing skills, as he mentioned, \textit{``Writing an email and then sending it to someone, making it a more formal tone or a more warm tone and then looking and seeing what words they added or changed was helpful in me learning how to write similar emails and stuff''}.
J1 also sees this feature as a learning opportunity, as she remarked, \textit{``I have a lot of difficulty with when writing an e-mail: asking questions that I think are sensitive or saying something I think is sensitive without expending anybody's energy. So I think there would definitely be a learning opportunity over time''}.}
One challenging aspect of the design is depending on the user's inputs to choose keywords to set the email tones.
Offering predefined tone options for users to select from, rather than solely relying on them to type their context, is recommended as it would be more helpful.

\TwoX{The most prominent concern raised by our participants about this design is the quality of the AI-generated output.
AI-driven text is too generic and lacks personalization.
Verifying the content by considering inputs from other stakeholders can be useful, however, it needs to make sure family members or surroundings possess the required skills to refine the AI-generated content.
As S2 mentioned, \textit{``I have these writing samples from students who use these AI-driven technologies like chatGPT. I found these sentences to be too generic and there is no personal touch. So when AI is generating email first, then family members' screen to possibly add some personal touches, we assume that person has experience of doing it because not all moms and dads can do the correction all of a sudden''}.}
\RThree{study 2 result is superficial, lacks depth}

\subsection{\TwoX{Socially supported job preparation}}

\TwoX{The AI-enabled collaborative resume-building is perceived as a useful feature due to its ability to automatically extract relevant topics that users should focus on.
As J1 remarked, \textit{``It would be nice to have this done by the system, providing topics I need to work on, the whole process will be a lot smoother''}.
As open-ended questions are often challenging to job seekers, E3 felt positive about the mock interview feature because it may help job seekers better prepare to answer, \textit{``Getting into the interview with those open-ended questions just kills people. So if they had some idea of the questions that may come up and how to answer them, it would be better than nothing''}.
The participants felt this can improve their communication skills, and become more comfortable with the interview process.
Practicing answers with a time constraint helps job seekers prepare answers more quickly.
However, S4 highlighted the fact interview is the most critical phase because of its randomized nature and there is no certainty regarding the questions asked, making it impossible for anyone to be fully prepared.
As she said, \textit{``Because no matter how much prep you do, it is still potentially random unless they do have the questions ahead of time, which may happen or not''}.
J1 conveyed doubt regarding the effectiveness of the feature as it does not offer back-and-forth conversation practice that mimics real-life scenarios.}

Participants also gave suggestions for improvement.
E1 thinks including alternative forms of commonly asked questions by rephrasing or making them more ND-friendly will be useful, \textit{``Here are some of the common ways that interviewers ask about the future career plan, which all mean the same thing''}. 
Adding remote access to resources like self-taught materials in the module for interview preparation is beneficial. 
J4 suggested adding a feature for practicing simulated conversations in interviews and providing feedback from social support in both text and voice forms could be highly effective.

\subsection{\TwoX{Peer-neurodivergent networking}}

\TwoX{Our last design Peer-neurodivergent networking was one of the most desired features perceived as valuable by every stakeholder group.
J1 shared that advice from neurodivergent mentors can be extremely helpful navigating in the job market,\textit{``I think it would be nice to have mentors in the general field who were neurodivergent, as they have different perspectives on how you feel about things, like difficulties with your job and stuff like that. You could get advice on that''}.
E3 believes this network will have a great impact on fostering leadership skills as well,\textit{``Whenever somebody volunteers as a mentor, they are also growing their leadership skills. So there are all kinds of benefits to everybody''}.
S5 liked the feature of showing individuals' professional details and how many people they are mentoring whereas existing networking sites highlight presenting only social information.}

\TwoX{Participants discussed a few suggestions, for instance, connecting the neurodiversity network including existing social networks would allow access to other peer networks outside the neurodivergent community.
This way individuals with autism can expand their social circle overcoming the fear of networking.
As E1 said, \textit{``A way to integrate this with LinkedIn so that someone is not necessarily disclosing their diagnosis but can tap into peer networks outside of the neurodivergent community. Because my neurodivergent students have a lot of apprehension about reaching out for networking and job openings''}.
Including details about a particular spectrum helps identify and compile a list of peers who share the same syndrome.
As E1 said, \textit{``If I could see a person is dyslexic, or has ADHD, Tourette's syndrome, etc. it helps when I want to talk to the person who has the same sorts of challenges I have''}.}

\section{\Three{Implications for Design} and Discussion}
\label{sec:implication}

Synthesizing our findings from Study 1 and Study 2, \ThreeC{we first discuss the insights that have implications to be transferred in the current design to enhance collaborative job-seeking experience for people with autism}. 
\ThreeB{Next, we explore how longitudinal system-level research, supported by cutting-edge intelligent technology, can yield beneficial features for prospective job-seekers with autism, expanding upon the four directions in the design space}.
Finally, we highlight noteworthy opinions that may not be directly related to the design of future tools but have social implications in job-seeking for people with autism.
\OneAC{discussion issues are generic, should be deepen}

\subsection{\ThreeC{Implications for Current Design}}

This subsection introduces the design directions that have close relations to improving the limitations of the current design. 
The directions are connected to Table~\ref{tab:S1results}, Rethinking search features. 

\ThreeX{
 \textbf{Rethinking current search criteria for job}:
We identified that the commonly used types of search criteria broadly adopted in current tools have a mismatch between the ideal search criteria that people with autism can leverage in seeking relevant positions.
The current search features lack adaptability and fail to address the needs of job seekers with autism mainly because existing filters focus on conventional job parameters like job title, location, and salary--which aligns with past study for job seekers from marginalized backgrounds~\cite{10.1145/3476065}. 
Search parameters tailored to the needs of job seekers with autism should consider more soft skill-based aspects such as team size, sensory considerations, frequency of communications, the level of independence within job roles, sensory environment, and so forth as these elements significantly impact the comfort and productivity of people with autism in their work environments.
However, this also requires a more active involvement of employers in providing detailed information about their work environment which potentially can pose challenges and uncertainty.
One common misconception among employers is the belief that accommodating individuals with autism is financially costly. 
Our Expert participants can play an active role in dispelling such misconceptions by promoting awareness regarding the significance of supporting autism in the workplace. 
Encouraging employers to establish a supportive work environment for employees to achieve their best performance, including adjustments in lighting, noise control, or increased empathy from co-workers which are not costly to implement.
Experts can advocate for ND training for employers and recruiters, stressing the need for an inclusive hiring process that values the perspectives and talents of job seekers~\cite{ara2023exploring}.}
\TwoAC{In discussion, discuss the context of design. Don’t reiterate related work}

\ThreeX{
\textbf{Meet accessibility guideline}~\cite{W3g} provide a unified approach for designing web content, offering accessibility features that can accommodate various disabilities.
To effectively cater to the diverse needs of individuals with autism, it is essential to integrate the principles outlined in this accessibility guidelines.
Particularly, avoiding ableist language and refraining from using misunderstanding terms.
Our ND participants have highlighted the importance of flexibility in color and font choices within designs, as these aspects greatly impact their usability. 
By following these guidelines and employing explicit visual cues such as neurodivergent-friendly fonts, icons can enhance accessibility, making the tool more accommodating for these end-users.}

\ThreeX{
\textbf{Support personalized multi-modal notifications}
To address the alarm fatigue issue, it is beneficial to create multi-modal reminders, incorporating audio and visual formats that can be customized based on personal preferences. 
For instance, autistic job seekers could set alarms using their preferred sounds, video clips, or pop-up animations which would easily be recognizable by them and grab their attention immediately. 
Using repetitive or monotonous tones as reminders might fail to capture their attention, leading to missing important notifications.
As they also have specific sensitivities, providing options to adjust colors, patterns, brightness, and volume will also be helpful to customize the alarm accordingly.}

\subsection{\ThreeB{Implications for Future Design}}

\ThreeX{In the design space, we derived a series of group tasks that could help people with autism engage in collaborative job-seeking. 
This section elaborates on how domains of CSCW, Human-AI Collaboration, computational machine learning, and assistive computing can use their underlying research problems to make progress in supporting the tasks.}

\ThreeX{From the perspective of \textbf{Human-AI Collaboration}: job-seeking is one of the most cognitively intensive tasks that requires full attention and long focus.
It can easily exhaust the job-seekers, and unfortunately, people with autism may have more difficulties than NTs. 
With our participants, we discussed various ideas on how we can leverage AI to streamline essential steps in job searching to better support their job-related tasks.
By applying a set of AI-driven intelligent features, we believe that future design can offload the cognitive burden for people with autism, so that they can achieve the varying tasks, such as writing and reading e-mails, making applications, and preparing interviews, with less attention. 
In supporting such individual tasks, there can be several AI-driven automation and facilitation features, such as an LLM-based executive functioning manager who can help list up the tasks and recommend priority tasks, a generative-AI driven resume builder, a mock-up interview agent who can synchronously generate new questions and give real-time suggestions to people with autism, interactive reader agent that receives the job call and enable interactive Q \& A to reduce the reading amount for people with autism, and many more.
Although AI can provide a reasonable initial material to work on, adding humans-- including people with autism and their collaborators--in the loop can add a much-needed personal touch to tailor these responses more specifically to job seekers' needs as per current state-of-the-art in human-AI collaboration studies~\cite{yan2022flatmagic,10.1145/3373625.3418305}.}

\ThreeX{From the perspective of \textbf{CSCW and Groupwork}: collaboration in job-seeking can make the job-seeker achieve better outcomes (as per server studies indicate~\cite{Liu2014-xs, paul2009unemployment, schwarzer2007functional}), common challenges in groupwork are cost for communication~\cite{hong2019design} and group awareness~\cite{hong2018collaborative}. 
From the angle of collaborative job-seeking for people with autism, we found that career experts are rather scarce and many of our participants didn't have a chance to work with them.
When it comes to the group collaboration between people with autism and experts, future design may benefit by defining the common types of information exchanged in communication, such as which positions the seeker is working, which hiring stage the job seeker is at, what preparation materials are covered, and what tasks to do, and transfer this to efficient communication format, such as markers, visualizations, and present it. 
Meanwhile, leveraging intelligent features for group communication can help experts use less time to work on different tasks, helping them to support more people with autism than before in varying tasks, such as connecting their clients with potential employers, suggesting job-related how-to materials to their clients, giving feedback, helping clients to do mock-up interviews, managing the client's expectation in applying, helping them to be mentally sturdy, and many more.} 

\ThreeX{From the perspective of \textbf{Assistive Computing}: while we identified a series of challenges, one major challenge in the job-seeking scenario is communication differences. 
We think the development of an autism-friendly context derivation model that can provide a communication support feature can benefit future job seekers and their collaborators by large.
As we identified through studies, people with autism tend to be ``literal thinkers'' who may have differences in interpreting the context behind questions, such as ``Imagine where you see yourself in 3 years'' as they may not know if they are still working at the company 3 years later.
This explains why a translation feature can be beneficial, which can help people with autism to understand the context hidden in NT language.}

\subsection{\ThreeA{Discussion}}

\ThreeX{The challenges faced by individuals with autism in professional settings, particularly in job seeking, revolve around the concept of Social Camouflaging (also known as masking)~\cite{pearson2021conceptual,miller2021masking} —where concealing one's authentic autistic traits becomes a significant dilemma.
This decision carries potential positive and negative consequences, influenced by financial pressures, concerns about social stigma, and experiences of discrimination. 
Job seekers often feel compelled to mimic neurotypical behavior, causing emotional strain due to the uncertainty about revealing their neurodivergence impacting their job prospects.
The constant effort to suppress natural behaviors can lead to increased stress, anxiety, and mental health issues~\cite{pearson2021conceptual}. 
It also creates a sense of disconnection from one's true self and can contribute to feelings of isolation and alienation.
Unfortunately, there is no straightforward solution to this dilemma. 
Moreover, societal perceptions perpetuate ableism~\cite{ableism}, leading to biases and unfair treatment in career searches for those with autism.}
Individuals with autism are still doing the \textit{``heavy lifting''} (E2), excelling in their professions, and actively advocating for themselves and their community.
Ideally, this effort should come from all stakeholder groups.
Changing this perception is a long-term process, requiring efforts beyond overnight change.

From our studies, we identified \textbf{Personalization} and \textbf{Individualization} are key factors in ensuring
that technology effectively addresses the individual needs and preferences of users (J8). 
Simply designing technology “for” people with disabilities without their active participation can lead to solutions that fail to genuinely address their requirements.
Shifting from a “design for” approach to a “design with” approach can be effective, however, this issue can still persist if researchers choose individuals with autism to fit a pre-conceived use case, decide which questions to ask, and determine the relevance of insights~\cite{dongle}. 
The "Disability Dongle" concept emphasizes involving individuals with disabilities, like autism, in the technology design process to cater to their requirements~\cite {dongle}. 
It is essential to adopt a more empathetic and user-centric approach to remove the “social-technical gap”~\cite{ackerman2000intellectual} \ThreeX{while currently, new designs are prioritizing the needs of target stakeholder groups only.} 
\ThreeX{Motivated from this perspective, the objective of our Study 1 was solely to acquire a profound understanding of the existing challenges and practices of our stakeholder groups. 
Study 2 was structured to involve active engagement from participants during the design stages, aiming to prevent the resulting solution from becoming merely a disability dongle.}

\ThreeX{Through our study, we also came to know about specialized job-seeking platforms designed for neurodivergent individuals offering remarkable features that greatly benefit job seekers. 
These tools, like Mentra~\cite{WinNT3} and MindShift~\cite{mindshift}, offer tailored support, guidance, and matching capabilities for job seekers. 
Mentra, for instance, prompts users during account creation to identify necessary accommodations, while MindShift provides job coaching and HR expertise. 
Setworks~\cite{serworks}, another tool, maintains comprehensive profiles and past interactions, aiding job seekers in their employment journey. 
There is still potential for enhancements to increase their accessibility and usefulness by incorporating the design considerations outlined in this work, fostering a collaborative approach to job searching that includes social support.}
\ThreeX{Aside from designing ND-friendly tools, there is a pressing need to shift employer perspectives towards valuing aptitude and potential over mere skill sets during interviews. 
Job seekers emphasize the importance of initiatives of new tools that challenge stereotypes and foster a more inclusive job market.
Individuals with autism often excel in recognizing ambiguous language and communicating precisely, which can be a significant asset in the workplace. 
This skill enables them to identify and address potential issues before they escalate, particularly in situations where neurotypical individuals may assume a universal understanding of communications, such as in legal terms, safety-critical systems, or customer commitments, which can have significant financial, health, and safety implications (J2).
However, our understanding is \textbf{Technology alone can not “fix” everything} because the societal stigma surrounding autism is a complex issue deeply ingrained in societal perceptions and attitudes. 
Attempting to solely address this through a single design or interface may prove insufficient due to the multifaceted nature of this stigma. 
A comprehensive strategy might involve various technical designs, educational campaigns, media representation, and community engagement to foster broader awareness among neurotypicals. 
Encouraging large companies to engage with ND peer groups, and offering mentorship or support beyond active hiring phases, further contributes to a more supportive and inclusive employment landscape.}

\section{ Conclusion}

We aimed to understand how people with autism collaborate with their social surroundings in handling common challenges in the job-seeking process in Study 1.
We also found how the technology and the hiring process can introduce additional challenges in their collaboration.
Through the study results, we identified four main directions of design.
Through Study 2, we understood the potential of the directions. 
We hope that our findings can motivate researchers and practitioners to build improved designs that can facilitate people with autism to collaborate with their support network and make improved outcomes in their job-related endeavors.

\begin{acks}
The authors express gratitude for generous support from the National Science Foundation, Future of Work Grant No.2026513. The research has been partially supported by the National Institute on Disability, Independent Living, and Rehabilitation Research (NIDILRR grant number 90DPGE0009). The authors are also thankful to Dr. Andrew Hunt (Carnegie Mellon University), Temple University, Melwood org., Frist Center for Autism and Innovation - Vanderbilt University, and Linktalent org. for their invaluable contribution to this paper. 
\end{acks}

\bibliographystyle{ACM-Reference-Format}
\bibliography{subsections/99_ref}

\end{document}